\documentclass[1p]{elsarticle}
\biboptions{sort&compress}
\usepackage{amsmath}
\usepackage{amssymb,amsthm}
\usepackage{graphicx}

\newcommand{\ket}[1]{\left | #1 \right \rangle}
\newcommand{\bra}[1]{\left \langle #1 \right |}

\newcommand{\beq}{\begin{equation}}
\newcommand{\eeq}{\end{equation}}
\newcommand{\beqa}{\begin{eqnarray}}
\newcommand{\eeqa}{\end{eqnarray}}
\newcommand{\openone}{{\bf 1}}

\newtheorem*{theorem}{Theorem}
\newtheorem*{lemma}{Lemma}

\journal{Annals of Physics}

\begin{document}
\title{{\bf Quantum computational capability of \\ a 2D valence bond solid phase}}
\author{Akimasa Miyake}
\ead{amiyake@perimeterinstitute.ca}
\address{%
Perimeter Institute for Theoretical Physics, \\
31 Caroline Street North, Waterloo Ontario, N2L 2Y5, Canada}
\date{September 17, 2010} 

\begin{abstract}
Quantum phases of naturally-occurring systems exhibit distinctive collective phenomena
as manifestation of their many-body correlations, in contrast to our persistent technological 
challenge to engineer at will such strong correlations artificially.
Here we show theoretically that quantum correlations exhibited in the two-dimensional (2D) valence 
bond solid phase of a quantum antiferromagnet, modeled by Affleck, Kennedy, Lieb, and Tasaki 
(AKLT) as a precursor of spin liquids and topological orders, 
are sufficiently complex yet structured enough to simulate universal quantum computation 
when every single spin can be measured individually.
This unveils that an intrinsic complexity of naturally-occurring 2D quantum systems ---
which has been a long-standing challenge for traditional computers ---
could be tamed as a computationally valuable resource, even if we are limited not to create 
newly entanglement during computation.
Our constructive protocol leverages a novel way to herald the correlations suitable for deterministic 
quantum computation through a random sampling, and may be extensible to other ground states of
various 2D valence bond phases beyond the AKLT state.

\end{abstract}


\maketitle

\section{Introduction}
\label{sec:intro}

Daily stories about a quantum computer have earned in our mind its image as super postmodern technology 
based on an artificial full control of quantum systems in the highest possible precision.
Most current approaches to implement a quantum computer are based on a bottom-up idea
in that we intend to build it by combining key elementary objects, as well summarized by
the celebrated DiVincenzo criteria \cite{divincenzo00}.
The idea implies that it should be equally, i.e., reversibly, doable to both create and annihilate  
at will many-body correlations (or entanglement if its nonlocal character has to be emphasized) among 
many qubits artificially. 
However, despite various promising candidates for a qubit and remarkable experimental progresses 
in their physical implementation, this {\em unitary} control of many-body entanglement in a scalable 
fashion is believed to remain the hardest challenge.

On the other hand, in a historical perspective, humanity has strived to find and tame resources present 
naturally on the earth, rather than achieving our goal from scratch by our immediate ability.
Speaking of energy resources for instance, instead of establishing self-contained energy cycle 
in the first place, we have learned to take advantage of more and more 
elaborated natural resources, e.g., from wood and water to oil and nuclear
(although their sustainable use is still our long-term project).
Here, based on the fact that nature realizes various quantum phases as manifestation of underlying 
many-body entanglement, we suggest taking a complementary, top-down vision in that we attempt to 
tame a resource of suitably structured entanglement, which could either exist in nature or be simulated
relatively naturally within our technology.

A key point is how to live with our limited ability. Once a specific natural resource of
structured many-body entanglement is provided, we are supposed to utilize only operations
which just consume entanglement without its new creation, such as local measurements
and local turning off of an interaction.
In this regard, our approach can be best compared to hewing out an arbitrary quantum evolving system
from the ``carving'' resource, in enhancing Feynman's original intuition 
\cite{feynman82} about quantum simulation in a computationally universal manner.
A perquisite is that we could be content with limited technology to prepare only one specific resource of
structured many-body correlation, that is less demanding than the full unitary control over arbitrary correlation. 
Moreover, if such a resource is available as a stable ground state of a naturally-occurring 
system, that gives us several possibilities to prepare, such as dissipative 
coolings or adiabatic evolutions, in a similar way that nature presumably does.
Our target of the naturally-occurring two-dimensional (2D) system is the valence bond solid (VBS) phase of 
spin $\tfrac{3}{2}$'s on the 2D hexagonal lattice, modeled by Affleck, Kennedy, Lieb, and Tasaki (AKLT) 
\cite{AKLT87,AKLT88}, which is widely recognized as a cornerstone in condensed matter physics.
Their VBS construction of the ground state in terms of the distributed spin singlets (or the valence 
bonds) has become one of most ubiquitous insights in quantum magnetism as well as in high-$T_c$ 
superconductivity, and leads to modern trends of spin liquids and topological orders.

It turns out here that the 2D VBS phase, represented by the AKLT ground state, provides
an ideal entanglement structure of quantum many-body systems that can be suitably {\it tamed}
through our limited capability to the goal of universal quantum computation.
Our top-down vision is materialized conveniently in taking advantage of a conventional 
framework of measurement-based quantum computation (MQC). Its methods have been developed
to ``steer'' quantum information through given many-body correlations using only a set of local 
measurements and classical communication, under which entanglement is just consumed without 
new creation. Later we extend that in a wider program to tame naturally-occurring many-body correlations.

\begin{figure}[t]
\begin{center}
\includegraphics[width=8.5 cm]{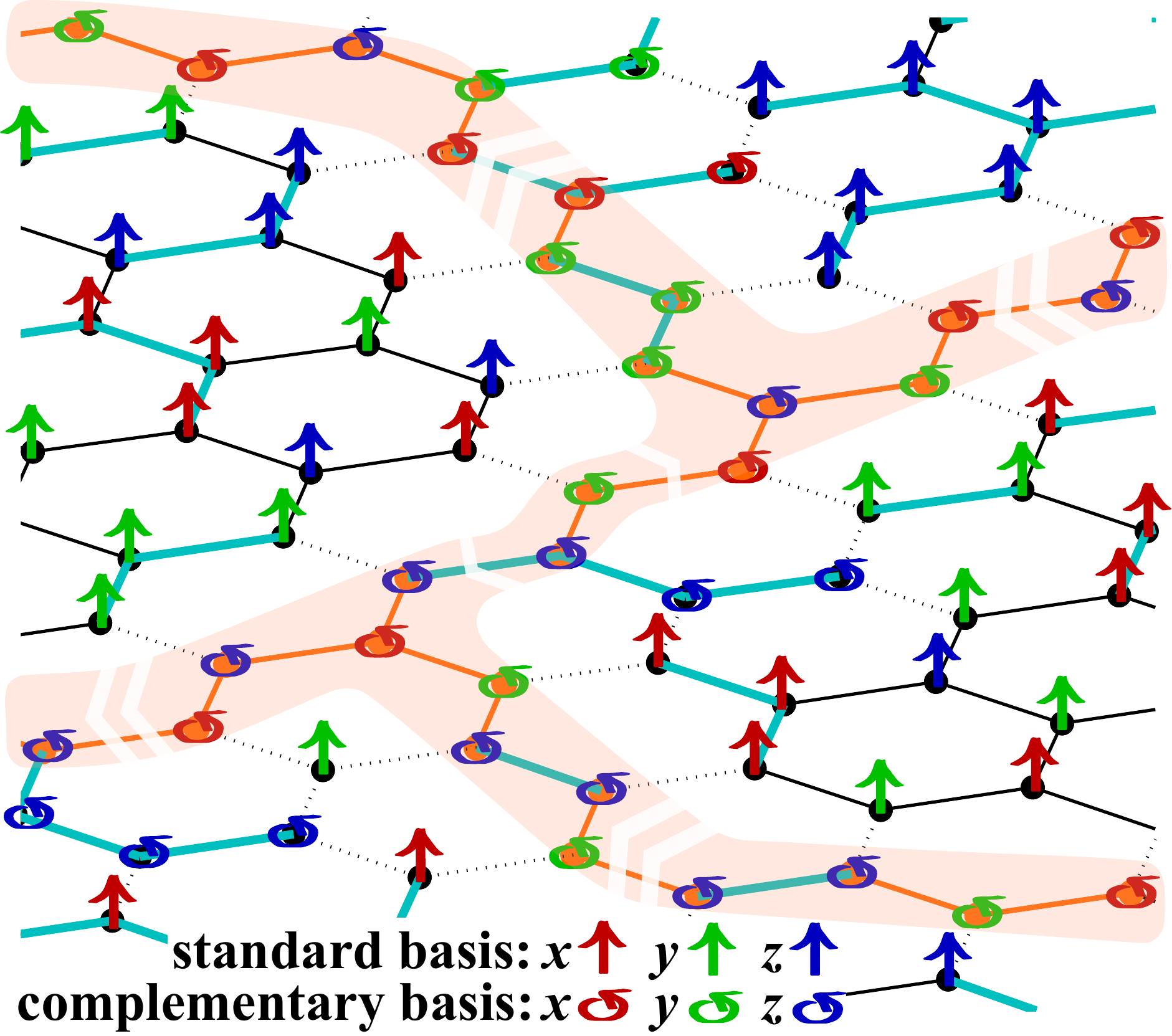} 
\end{center}
\caption{A scheme of quantum computation through measuring the correlations of the 2D AKLT state,
a representative state of the 2D VBS phase of spin $\tfrac{3}{2}$'s on the 2D hexagonal lattice.
After a random sampling which assigns every spin per site to one of three axes $x$, $y$, and $z$, 
the typical configuration of the outcomes enables us to choose the backbone structure (described by   
a shaded pinkish-orange region) along which quantum computation is deterministically simulated
in terms of a quantum circuit.
Our protocol harnesses a pair (depicted as a dotted bond of the hexagonal lattice) of neighboring sites 
where one is measured in a standard basis and the other is done in a complementary basis, 
to accommodate the desired structure of space-time along the region of the backbone. 
An emergence of the time is simulated if both two bits of information out of measurements per site 
are communicated to the same direction (as depicted as the double arrows), on the other hand, 
an emergence of the space is simulated if two bits of information are communicated to the opposite 
directions (as depicted as a pair of the single arrows pointing apart). 
The figure, reminiscent of the Feynman diagram for scattering of two elementary particles, illustrates 
a microscopic view of the Figure~\ref{fig:matchedbonds}. The two-qubit CNOT gate is implemented 
in the middle region between two quantum logical wires running from the right to the left.}
\label{fig:scheme}
\end{figure}

In the context of MQC, the so-called 2D cluster state \cite{briegel01} is the first and canonical instance 
of such an entangled state that pertains to a universal quantum computational capability when
every single qubit (spin $\tfrac{1}{2}$) is measured individually and the outcomes of the 
measurements are communicated classically \cite{raussendorf01,raussendorf03}. 
Remarkably, it was already noticed in Ref.~\cite{verstraete04} that MQC on the 2D cluster state utilizes
a structure of entanglement which is analogous to that of the aforementioned VBS state. 
Following such an observation, the tensor network states, as a class of efficiently classically 
parameterizable states in extending the VBS construction, has been used in 
Refs.~\cite{gross07prl, gross07pra} to construct resource states of MQC, 
where it was indicated that a certain set of the local matrices or tensors which describe the correlations 
(cf. Eq.~(\ref{groundstate}) in our case) can result in a quantum unital map through the single-site measurement. 
Notably, however, most known examples considered so far, including additionally those e.g., in 
Refs.~\cite{bartlett06,brennen08,gross08,griffin08,chen09,cai10}, are constructed to have 
such a convenient yet artificial property --- as often referred as one of peculiar properties of the 
correlations of the 2D cluster state --- that it is possible to decouple deterministically (by measurements 
of only neighboring sites) a 1D-chain structure that encodes the direction of a simulated time as 
a quantum logical wire of the quantum circuit model.
This peculiarity is said to be artifact of another less realistic feature of the 2D cluster state 
in that it cannot be the exact ground state of any two-body Hamiltonian of spin $\tfrac{1}{2}$'s \cite{nielsen06,nest08}. 
Thus one cannot expect such convenience for the correlations of a {\it genuine} 2D ground state of 
a naturally-occurring spin system.

The main result of our paper is summarized in the following (informal) theorem and illustrated 
in the Figure~\ref{fig:scheme}.
As elaborated in the text, we introduce a novel way to herald the correlations suitable for {\it deterministic}
quantum computation through a {\it random} sampling, to tame for the first time the genuine 2D 
naturally-occurring correlation. Otherwise it has natural tendency to split an incoming 
information into two outgoing information because of certain symmetric nature of the {\it three} directions 
at every site of the 2D hexagonal lattice.
This seems to be the reason why MQC on the 2D AKLT state has been an open question in a long 
time, although the AKLT state by the 1D {\it spin-1} chain was shown in Ref.~\cite{brennen08} 
to be capable of simulating a single quantum wire of MQC.

\begin{theorem} 
Universal quantum computation can be simulated through consuming monotonically entanglement 
provided as the 2D AKLT state $\ket{\mathcal G}$ (defined as the VBS state by a spin $\tfrac{3}{2}$ per site 
and described as a tensor network state of Eq.~(\ref{groundstate})) of the size proportional to the target 
quantum circuit size.
To this end, we leverage single-site measurements of every individual spin $\tfrac{3}{2}$, a bounded amount of 
classical communication of measurement outcomes per site, and efficient classical side-computation.
\end{theorem}

We note that upon completion of the work, another approach to suggest computational usefulness of
the 2D AKLT state is presented independently in Ref.~\cite{wei11} by Wei, Affleck, and Raussendorf recently, 
through transforming the 2D AKLT state into a 2D cluster state using a clever mastery of 
the graph states (an extension of the cluster state to a general graph).
In comparison, our proof constitutes a more direct protocol to use the correlations 
of the 2D AKLT state to simulate straightforwardly the quantum circuit model, 
with much less resort to the known machinery of the cluster state.
Our construction of the protocol also respects more explicitly the physics and topological 
nature of the 2D VBS phase, 
such as the edge states at the boundary and the (widely-believed) energy gap at the bulk.
It clarifies their {\it operational} usage in quantum information in that
the former is used to process the logical information of quantum computation in the degenerate
ground subspace and the latter could contribute to its protection, in providing some robustness 
against local noise, as the ground-code version of MQC discussed in the Section~\ref{sec:conclude}.

\section{2D valence bond solid ground state}
\label{sec:vbs}

The 2D VBS phase can be modeled by a nearest-neighboring two-body Hamiltonian of the
antiferromagnetic Heisenberg-type isotropic interaction (i.e. $J > 0$) \cite{AKLT88,KLT88}, 
\beq
H = J \sum_{(k, k')}^{\rm n.n.} \left[{\mathbf S}_{k} \cdot {\mathbf S}_{k'} + 
\frac{116}{243} ({\mathbf S}_{k} \cdot {\mathbf S}_{k'})^2 + 
\frac{16}{243} ({\mathbf S}_{k} \cdot {\mathbf S}_{k'})^3 \right] ,
\label{H}
\eeq
where ${\mathbf S}_k$ is the spin-$\tfrac{3}{2}$ irreducible representation of $\mathfrak{su} (2)$ at the site $k$,
and the summation is taken over all the nearest neighboring pairs $(k, k')$ of spin $\tfrac{3}{2}$'s on 
the 2D hexagonal lattice.
The particular weights to the biquadratic and bicubic terms are chosen conventionally to be the projector onto
the subspace of the total spin $3$ for every pair of $(k, k')$. However the 2D VBS phase itself is supposed to persist
around this AKLT point without a fine tuning of these weights, in the same way as the 1D case.
It is important to mention that our 2D VBS phase should be distinguished from the 2D valence bond 
crystal (VBC) phase, since VBC is usually used to refer to the phase that consists of the valence bonds
in a broader sense, namely including not only the VBS phase but also the dimer phase etc.
However, there are considerable differences between VBS and the dimer phase, for example, 
regarding the global nature of entanglement and the origin of the ground-state degeneracy (with 
an open boundary condition).

The 2D AKLT ground state is such a VBS wavefunction that the symmetrization of three (virtual) spin $\tfrac{1}{2}$'s 
to represent a physical spin $\tfrac{3}{2}$ per site is made on a collection of the singlet pairs 
$\tfrac{1}{\sqrt{2}}(\ket{1^z}\otimes\ket{0^z} - \ket{0^z} \otimes \ket{1^z})$ of spin $\tfrac{1}{2}$'s, 
where each singlet is distributed along every bond of the 2D hexagonal lattice.
The construction can be visualized like in the Figure 3.2 of Ref.~\cite{AKLT88} for instance.
It is straightforward, for our convenience, to describe it as a tensor network state 
via the celebrated Schwinger boson method
\cite{arovas88}, 
\beq
\ket{\mathcal G} = \sum_{\alpha_k, \alpha_{k'}} {\rm tr} 
\left[B \prod_{k \in \top}  A_{\top}[\alpha_k]  \ket{\alpha_k} \prod_{k' \in \bot} A_{\bot}[\alpha_{k'}] \ket{\alpha_{k'}}\right] , 
\label{groundstate}
\eeq
where $\alpha_{k (k')}$ at the site $k$ (or $k'$) runs over
$\tfrac{3}{2}^{z},\tfrac{1}{2}^{z}, -\tfrac{1}{2}^{z}, -\tfrac{3}{2}^{z}$, and the trace is taken by the contraction of the tensors according to their locations on the 2D hexagonal lattice. 
The boundary condition is assumed to be open in that we are simply given a finite bulk portion of the lattice,
and, the boundary tensor $B$ is set to be the identity according the~\ref{app:boundary}.
The tensors at the site with the $\top$-shaped or $\bot$-shaped bonds are found to be given by 
\begin{align}
\begin{split}
A_{\top}[\tfrac{3}{2}^z] &= |0^{z}\rangle\langle 1^{z}| \otimes \bra{1^z} , \\
A_{\top}[\tfrac{1}{2}^z] &= \frac{-1}{\sqrt{3}} \left(|0^{z} \rangle\langle 1^{z}|\otimes \bra{0^z} + Z \otimes \bra{1^z}\right) , \\
A_{\top}[-\tfrac{1}{2}^z] & =  \frac{1}{\sqrt{3}} \left(-|1^{z}\rangle \langle 0^{z}|\otimes \bra{1^z} + Z \otimes \bra{0^z}\right) , \\
A_{\top}[-\tfrac{3}{2}^z] & = |1^{z}\rangle\langle 0^{z}| \otimes \bra{0^z} ,
\end{split}
\label{Atop}
\end{align}
\begin{align}
\begin{split}
A_{\bot}[\tfrac{3}{2}^z] &= - |0^{z}\rangle\langle 1^{z}| \otimes \ket{0^z} , \\
A_{\bot}[\tfrac{1}{2}^z] &= \frac{1}{\sqrt{3}} \left(- |0^{z}\rangle\langle 1^{z}|\otimes \ket{1^z} + Z \otimes \ket{0^z}\right) , \\
A_{\bot}[-\tfrac{1}{2}^z] & =  \frac{1}{\sqrt{3}} \left(|1^{z}\rangle\langle 0^{z}|\otimes \ket{0^z} + Z \otimes \ket{1^z}\right) , \\
A_{\bot}[-\tfrac{3}{2}^z] & = |1^{z}\rangle\langle 0^{z}| \otimes \ket{1^z} ,
\end{split}
\label{Abot}
\end{align}
respectively. Here the first ket and bra correspond to degrees of freedom by the left and the right,
and the second ket or bra corresponds to that by the up or the down, respectively.
The Pauli matrices are defined as 
$Z = |0^z\rangle\langle 0^z | -|1^z\rangle\langle 1^z | $,
$X = |0^z\rangle\langle 1^z | +|1^z\rangle\langle 0^z | $, and $Y = i XZ$ with the imaginary unit $i = \sqrt{-1}$.
An element of the Pauli group, including the identity, will be denoted as $\Upsilon$ later.
In this tensor network description, the effective spin $\tfrac{1}{2}$ labeled by $\ket{0^z}, \ket{1^z}$
should better be understood as a manifestation of the fractionalized degree of freedom, the edge state, 
emergent at the boundary across a single bond.
Later it will turn out that 
these tensors are interpreted as the {\it logical} action on these emergent edge states (in other
words, degenerate ground states), when the spin is measured in the direction by its argument.

The set of tensors in terms of the other bases has exactly the same 
structure (up to a possible {\it overall} phase) because of the rotational symmetry.
In defining $\ket{0/1^x} = \tfrac{1}{\sqrt{2}}(\ket{0^z} \pm \ket{1^z})$ and 
$\ket{0/1^y} = \tfrac{1}{\sqrt{2}}(\ket{0^z} \pm i \ket{1^z})$,
\begin{align}
\begin{split}
& A_{\top}[\alpha^x] = A_{\top}[\alpha^z] |_{z \mapsto x}, \quad A_{\bot}[\alpha^x] = - A_{\bot}[\alpha^z] |_{z \mapsto x} , \\
& A_{\top}[\alpha^y] = -i A_{\top}[\alpha^z] |_{z \mapsto y}, \quad A_{\bot}[\alpha^y] = A_{\bot}[\alpha^z] |_{z \mapsto y} , 
\end{split}
\label{Axy}
\end{align}
where $\alpha = \tfrac{3}{2},\tfrac{1}{2}, -\tfrac{1}{2}, -\tfrac{3}{2}$ and
our notation, for example for the $S^x$ basis, is meant to describe the tensors obtained in 
replacing $\ket{0/1^z}$ and $\bra{0/1^z}$ into $\ket{0/1^x}$ and $\bra{0/1^x}$ as well as $Z$ into $X$
in Eqs.~(\ref{Atop}) and (\ref{Abot}).

The 2D AKLT state inherits various characteristics from the 1D VBS state. For example, 
its correlations measured in terms of the two-point function was shown to decay
exponentially in lattice distance with the correlation length $\xi = 1/\ln (3/2) \approx 2.47 $ \cite{KLT88}.
Furthermore, in very contrast with the dimer phase, it also realizes a fractionalized degree of freedom on 
every boundary, called the edge state, as mentioned above.
A numerical calculation \cite{katsura10} of the entanglement entropy confirms a qualitative nature of edge states.
On the other hand, the spectral gap to the excited states in the 2D AKLT model is widely believed to persist 
in the thermodynamical limit, but has yet to be proved.

\section{Insight to the MQC protocol}
\label{sec:outline}

\begin{figure}[t]
\begin{center}
\includegraphics[height=0.48\columnwidth]{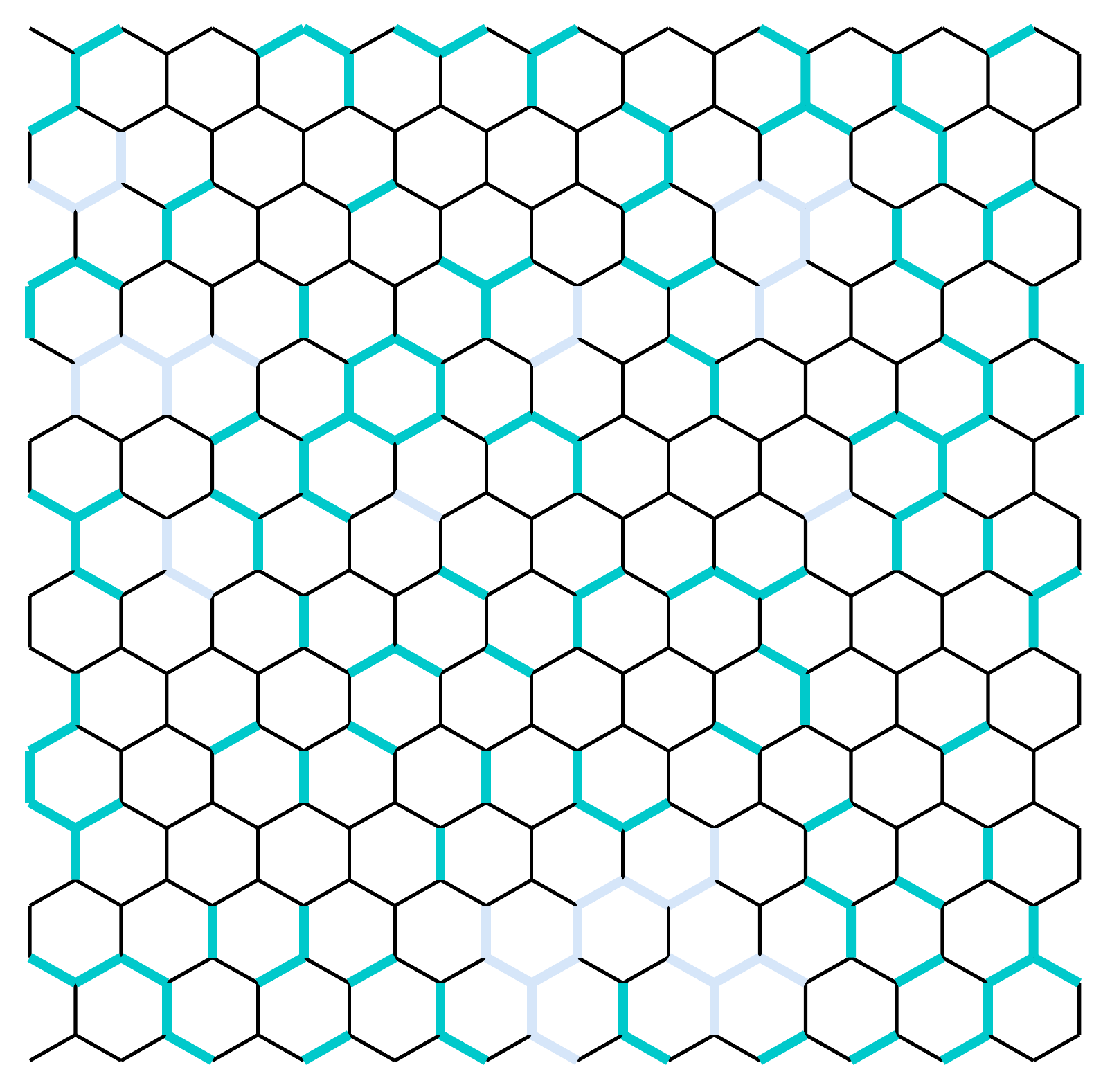}  
\includegraphics[height=0.48\columnwidth]{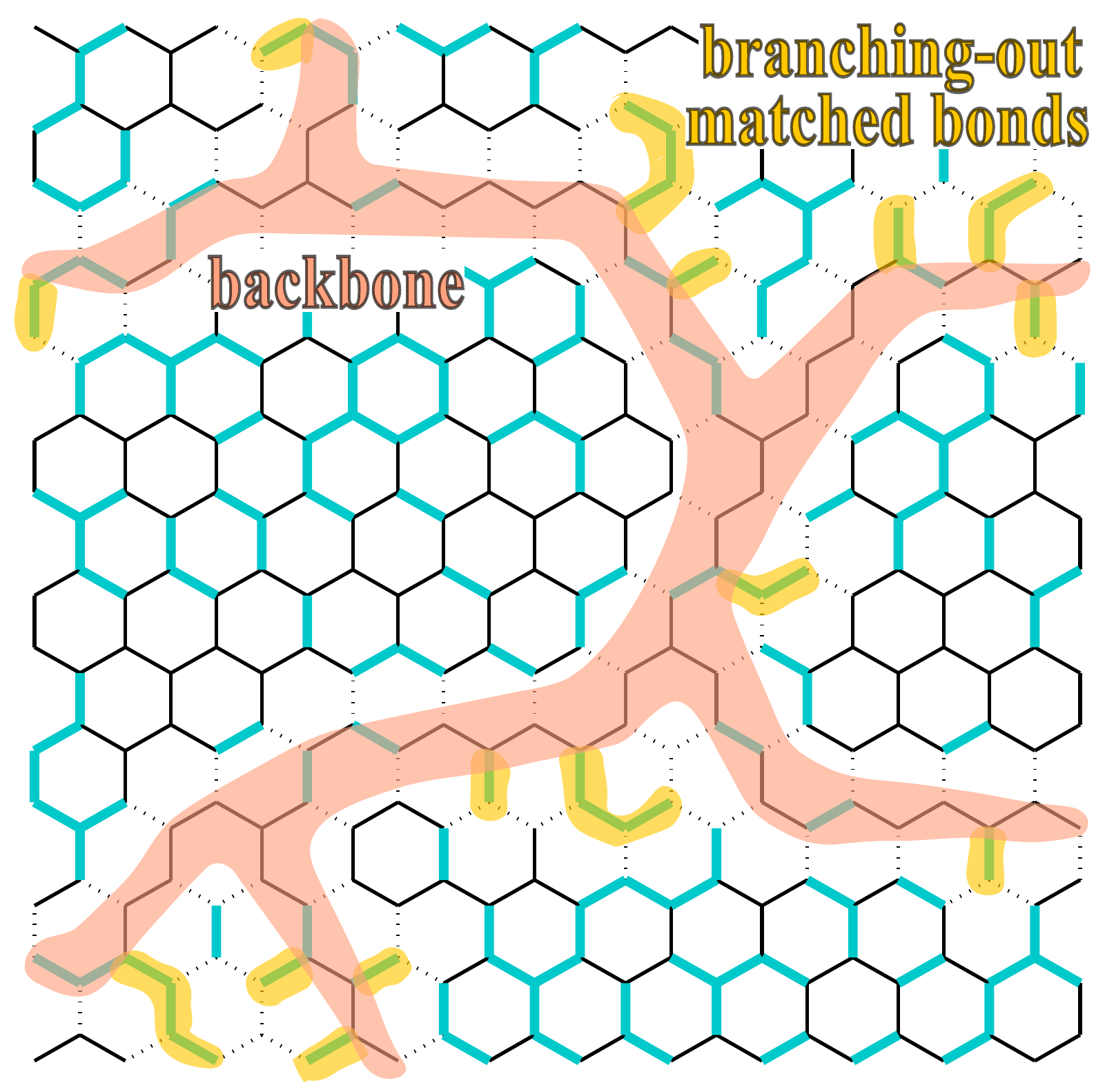}  
\end{center}
\caption{(Left) Illustrated is a typical distribution of matched bonds depicted as thicker (dark-cyan 
and light-purple) bonds. Here the bond is called matched when the pair of neighboring sites is 
assigned to the same axis through the polarizing measurement at the first stage of the protocol. 
The region to simulate a quantum circuit by the complementary bases, called the backbone,  
is going to be chosen by classical side-computation, so as to guarantee its global topology is identical 
to that of the target quantum circuit (cf. the Section~\ref{sec:global}).
To this end, the backbone must skirt some ``off-limits'' configurations of matched-bond clusters
that induce additional undesirable logical loops if all of them are involved in the backbone.
Accordingly, it is sufficient that every quantum logical wire is chosen from any horizontally percolating 
path that circumvents only some clusters of matched bonds, for instance, colored in light purple.
(Right) A possible embedding of the backbone, described as a shaded pinkish-orange region, has been 
identified by analyzing classically the occurrence of matched bonds .
The ``branching-out'' matched bonds that are involved in the backbone and described by a shaded yellow region
are the extended region where the logical information of quantum computation is processed in the second stage of the
protocol. However, by having no additional closed loop, they would not disturb the desired logical action.
Based on the statistical property originated from the genuine 2D nature of correlations of the AKLT state,
the almost sure success of such an identification of the suitable backbone is guaranteed 
in an analogous way with the emergence of the macroscopic percolating cluster in the bond percolation phenomenon.
The microscopic view near the CNOT gate is highlighted in the Figure~\ref{fig:scheme}.}
\label{fig:matchedbonds}
\end{figure}

We intend to simulate the quantum circuit model through measuring the correlations at every site.
We call the part of the 2D hexagonal lattice sites 
that corresponds to the quantum circuit (consisting of the quantum logical wires running almost horizontally 
and their entangling gates described vertically) a {\it backbone}, as seen in the Figure~\ref{fig:matchedbonds}.
The degree of the backbone site refers to how many neighbors it has along the backbone. 
The degree-3 backbone sites are used at every junction of the horizontal logical wire
with a vertical entangling gate, so that they are required only occasionally.

A key insight to construct our protocol
is that since the reduced density operator of every spin $\tfrac{3}{2}$ per site is totally mixed 
and isotropic, i.e., the normalized identity projector $\tfrac{\openone}{4}$, we are able to extract
{\it 2 bits} of classical information by measurements per site. 
Then it is sensible that in stead of obtaining them at once, a part of the information, indeed $\log_2 3$ bits 
in our case, is first extracted and we adapt the next stage according to it.
It might be surprising that the first measurement induces a kind of randomization, but intuitively speaking,
this part is crucial to separate the original quantum correlation that intrinsically involves genuine 2D fluctuations 
into the classical correlation (or, a statistical sampling) that can be still efficiently handled by a classical 
side-processor and the ``more rigid'' quantum correlation suitable for {\it deterministic} quantum computation. 
A global statistical nature of the AKLT correlations through the first stage guarantees, in an analogous way with 
the classical percolation phenomenon, that an embedding of the backbone (i.e. the target quantum circuit) can 
be found in the {\it typical} configuration of a heralded, randomized distribution of entanglement.
At the second stage which implements quantum computation, the protocol
is invented in such a way that the standard-basis measurement and complementary-basis one, 
both of which are defined later, are paired (as depicted by the dotted bonds in the Figure~\ref{fig:scheme}).
This paired processing is crucial to make the logical information bound in the domain 
of the backbone, somehow in a reminiscent way how the deoxyribonucleic acid (DNA) holds genetic
information in its famous double strands.

\subsection*{Summary of the MQC protocol}

Now we outline our MQC protocol, which consists of two stages.
(i) The first stage is to apply a measurement $\{M^{x},M^{y},M^{z}\}$ which polarizes randomly 
toward one of the three orthogonal axes at every site. 
We define a degenerate projection $M^{\mu} \;(\mu = x,y,z)$ as
\beq
M^{\mu} = \sqrt{\tfrac{2}{3}}(|\tfrac{3}{2}^\mu\rangle\langle \tfrac{3}{2}^\mu |+ 
|-\tfrac{3}{2}^\mu \rangle\langle -\tfrac{3}{2}^\mu|).
\eeq
The set of $\{M^{x}, M^{y}, M^{z} \}$ constitutes the positive operator value measure (POVM)
by satisfying $\sum_{\mu=x,y,z} M^{\mu \dagger} M^{\mu} = \openone $, 
so that it is a valid local measurement with three alternative, random outcomes $\mu$.
We must record the outcome $\mu_{k} $ at every site $k$ and 
collect the location of such {\it matched} bonds that the pair of the axes 
for the neighboring sites $k, k' $ coincides, namely $\mu_k = \mu_{k'}$ .
Based on the occurrence of matched bonds (which need additional care in their use),
we are able to determine the backbone by efficient classical side-computation in circumventing
some rare ``off-limits'' configurations of matched bonds defined later.

(ii) The second stage carries actual quantum computation, using further projective 
measurements at every site and feedforward of their outcomes.
Once the backbone is identified, the computation is deterministic in a very similar way with MQC
on the 2D cluster state.

\section{Widgets for the logical gates}
\label{sec:gates}

A basic idea of the widgets to construct the logical gates from the set of tensors is outlined first.
Suppose the bond is not matched for a pair of the nearest neighboring sites,
then either of the sites can be used as the part of the backbone, by treating
the other as the non-backbone site. 
Let the assignment of the first-stage polarizing measurement at the backbone site
be $\mu$ and the other be $\nu (\ne \mu)$, and, as an illustration, assume $\mu = z$ at the $\top$ site 
and $\nu = x$ at the $\bot$ site.
It turns out that, in measuring the non-backbone site in the basis of $\{\bra{\tfrac{3}{2}^{x}} ,\bra{-\tfrac{3}{2}^{x}} \}$ 
(since the state is already in this subspace by the preceding $M^{x}$), we can always get a unitary action 
at the backbone site along the other remaining two bonds, regardless of which outcome has occurred.
This is because conditioned on the specification of the tensor of Eqs.~(\ref{Abot},\ref{Axy}) to be, for instance, 
$|0^x\rangle\langle 1^{x} | \otimes \ket{0^x}$ by the outcome $\bra{\tfrac{3}{2}^{x}}$  at the non-backbone site, 
the tensors at the backbone site are essentially fixed as
\begin{align}
\begin{split}
\tilde{A}_{\top}[\tfrac{3}{2}^z] &= |0^{z}\rangle\langle 1^{z}| \otimes \langle 1^z | 0^x \rangle 
= \tfrac{1}{\sqrt{2}} |0^{z}\rangle\langle 1^{z}| , \\
\tilde{A}_{\top}[-\tfrac{3}{2}^z] & = |1^{z}\rangle\langle 0^{z}| \otimes \langle 0^z | 0^x\rangle 
= \tfrac{1}{\sqrt{2}}  |1^{z}\rangle\langle 0^{z}| .
\label{zidentity}
\end{split}
\end{align}
The overall numerical factor $\tfrac{1}{\sqrt{2}}$ implies the fact that the outcome $\bra{\tfrac{3}{2}^{z}}$ would occur with
probability $\tfrac{1}{2}$, and of course the other outcome would do with the same probability $\tfrac{1}{2}$.
Indeed, the measurements used for the unitary actions are unbiased in their output probability.
Here after, however, we do not describe the overall factor explicitly, since it is irrelevant to the statistics of 
the logical output of quantum computation.
It can be readily seen that we can get a unitary gate by mixing these two tensors
in such a way that for instance, measuring this backbone site in the basis 
$\{\tfrac{1}{\sqrt{2}}(\bra{\tfrac{3}{2}^z} + \bra{-\tfrac{3}{2}^z}), \tfrac{1}{\sqrt{2}}(-\bra{\tfrac{3}{2}^z} + \bra{-\tfrac{3}{2}^z}) \}$
provides $X (= \tilde{A}_{\top}[\tfrac{3}{2}^z] + \tilde{A}_{\top}[-\tfrac{3}{2}^z])$ or 
$XZ (= -\tilde{A}_{\top}[\tfrac{3}{2}^z] + \tilde{A}_{\top}[-\tfrac{3}{2}^z] )$ respectively.
Both unitary actions should be interpreted as the logical identity 
since the difference by a Pauli operator can be readily incorporated, as explained later, by the adaptation of
the following measurement basis so that the Pauli operators are treated as the byproduct 
operator $\Upsilon$.
In more general, measuring in a basis on the plane spanned by the aforementioned two vectors is found to
provide the logical rotation $R^{z} (\theta) = |0^z \rangle \langle 0^z | + e^{i \theta} |1^z \rangle\langle 1^z |$ 
by an arbitrary angle $\theta$ along the axis $z$.
We call any of such bases {\it complementary} to the standard basis 
$\{\bra{\tfrac{3}{2}^{\mu}}, \bra{-\frac{3}{2}^{\mu}} \}$ given the axis $\mu$ at the site.
The choice of the complementary basis depends on the axis $\nu$ of the associated non-backbone site.
 Accordingly, we may denote the complementary basis in an abstract way as 
$\{\bra{\gamma^{\mu | \nu}} \}$, which will be explicitly defined later.

Such a construction of a ``mutually-unbiased'' input to the backbone is the key technique
provided by the pair of the standard-basis and complementary-basis measurements.
Indeed, this basic idea is shown to be extensible with an additional care even when the bond is matched.
As it will get clearer, the set of sites measured in the complementary bases, which covers strictly the backbone,
corresponds to the region where logical information of quantum computation is processed.
Our construction enables us to decouple the logical output probability of quantum computation from the rest 
of the sites measured in the standard-basis measurements (conditioned on the configuration of the axis per site
after the first stage of the protocol).
A suitable identification of the backbone is addressed later in the Section~\ref{sec:global} as a global nature
of computational capability, supported by a prescription to the matched bonds in the~\ref{app:match}.

\subsection{Single-qubit logical gates}

Any single-qubit logical gate in $SU(2)$ can be implemented by the sequence of the rotations
$R^{\mu} (\theta) = |0^\mu\rangle\langle 0^\mu| + e^{i \theta} |1^\mu \rangle\langle 1^\mu|$ 
along two independent axes, using the Euler angles $\theta$'s.
Here we take $\mu = z$ and $x$ without loss of generality, and describe the detailed protocol for
$R^z (\theta)$.

The $R^z (\theta)$ can be only attempted if the axis $\mu$ of the backbone site is assigned to be 
$z$ after the polarizing measurement at the first stage. Otherwise we teleport the logical information
along the backbone by implementing the logical identity by measuring the site in the complementary
basis with the fiducial angle (i.e., $\theta = 0$) until we reach the backbone site with $\mu = z$. 
Let the axis of the associated non-backbone site be $\nu$, and the outcome when measured in 
the standard basis denoted as $c \in \{0,1\}$ in corresponding to $\bra{(-1)^c \tfrac{3}{2}^{\nu}}$.
The complementary basis for the given axis $\mu = z$ is defined (at both the $\top$ and $\bot$ sites) 
as, if $\nu = x$, 
\begin{equation}
\langle \gamma^{z | x} (\theta)| = \tfrac{1}{2\sqrt{2}}\left[ (1 + (-1)^{b} e^{i \theta})
\left(\bra{\tfrac{3}{2}^z} + \bra{-\tfrac{3}{2}^z}\right) + 
 (1 - (-1)^{b} e^{i \theta}) \left(-\bra{\tfrac{3}{2}^z} + \bra{-\tfrac{3}{2}^z}\right) \right] ,
\end{equation}
where $b \in \{0,1\}$ corresponds to each outcome of the orthogonal projective measurement 
with the angle $\theta$ or $\theta + \pi$ respectively, and if $\nu = y$,
\begin{equation}
\langle\gamma^{z | y} (\theta)| = \tfrac{1}{2\sqrt{2}}\left[ (1 + (-1)^{b} e^{i \theta})
\left(-i \bra{\tfrac{3}{2}^z} + \bra{-\tfrac{3}{2}^z}\right) + 
(1 - (-1)^{b} e^{i \theta}) \left(i\bra{\tfrac{3}{2}^z} + \bra{-\tfrac{3}{2}^z}\right) \right] .
\end{equation}
In all the cases, the logical action is given by
\beq
\sum_{\alpha} \tilde{A}[\alpha] \langle\gamma^{z|\nu} (\theta) | M^{z} | \alpha \rangle  = 
X Z^{b \oplus c} R^z (\theta), 
\label{Rz}
\eeq
where a partial contraction has been made, in the same way as in Eq.~(\ref{zidentity}), with 
the rank-1 tensor from the associated non-backbone site, using the notation
\beq
\tilde{A}_{\top}[\alpha] = A_{\top}[\alpha] \ket{c^{\nu}}, 
\quad
\tilde{A}_{\bot}[\alpha] = \bra{c \oplus 1 ^{\nu}}A_{\bot}[\alpha].
\label{tildeA}
\eeq
This is the desired rotation up to a Pauli byproduct $\Upsilon = X Z^{b \oplus c}$.
Note that the choice of the various complementary bases only depends on the axis $\nu$
and not on the outcome $c$ at the non-backbone site, so that once the axis of every site
has been assigned and the backbone structure has been determined, the backbone site
and its associated non-backbone site can be measured in a parallel way.
Likewise, the counterpart for $R^x (\theta)$ can be obtained by
\begin{align}
& \langle\gamma^{x | z} (\theta)| = \tfrac{1}{2\sqrt{2}}\left[ (1 + (-1)^{b} e^{i \theta})
\left(\bra{\tfrac{3}{2}^x} + \bra{-\tfrac{3}{2}^x}\right) + 
(1 - (-1)^{b} e^{i \theta}) \left(-\bra{\tfrac{3}{2}^x} + \bra{-\tfrac{3}{2}^x}\right) \right] , \\
& \langle\gamma^{x | y} (\theta)| = \tfrac{1}{2\sqrt{2}}\left[ (1 + (-1)^{b} e^{i \theta})
\left(i \bra{\tfrac{3}{2}^x} + \bra{-\tfrac{3}{2}^x}\right) + 
(1 - (-1)^{b} e^{i \theta}) \left(-i\bra{\tfrac{3}{2}^x} + \bra{-\tfrac{3}{2}^x}\right) \right] .
\end{align}
The measurement in these complementary bases result in $\Upsilon R^{x} (\theta)$ with 
$\Upsilon =  Z X^{b \oplus c}$.
Although it is straightforward to define $\bra{\gamma^{y|\nu} (\theta)}$ for $R^{y} (\theta)$, 
it is theoretically sufficient that the sites labelled by $\mu = y$ are
only used for the logical identity (with $\theta = 0$), whose dependence to the outcomes is
$\Upsilon = X^{b \oplus c} Z^{b \oplus c \oplus 1}$.
The dependence of the byproduct operator $\Upsilon$ on measurement outcomes is summarized in the 
Table~\ref{byproduct}.

In the above, we have assumed that the (degree-2) backbone site 
has in tow its neighboring non-backbone site which is measured in the standard basis along $\nu$
and provides the outcome $c$.
If the backbone site is connected {\it via the matched bonds} to its associated non-backbone site, then
the intermediate sites should be measured in the complementary basis too, and a correction to 
$\ket{c^{\nu}}$ (or $\bra{c \oplus 1^{\nu}}$) of $\tilde{A}[\alpha]$ defined by Eq.~(\ref{tildeA}) is required.
Roughly speaking, the non-backbone sites connected by the ``branching-out'' matched bonds (colored in yellow 
in the Figure~\ref{fig:matchedbonds}) need to be treated in the same way as the backbone sites, so that we have 
to sum up all their byproduct operators.
The motivation as well as technical detail of this prescription to the matched bonds is found in the~\ref{app:match}.

\begin{table}[t]
\begin{center}
{\tabcolsep = 0.4cm
\begin{tabular}{c|c|c}
\hline  \hline 
$\mu$  &  $a^x$  & $a^z$  \\
\hline
$\gamma^{x | \nu} $  & $b \oplus c$  & 1 \\
$\gamma^{y | \nu} $  & $b \oplus c$  & $b \oplus c \oplus 1$ \\
$\gamma^{z | \nu} $  & 1  & $b \oplus c$ \\
\hline \hline
\end{tabular}}
\end{center}
\caption{The indices $(a^x , a^z)$ of the byproduct operator $\Upsilon = X^{a^x} Z^{a^z}$ for the complementary 
basis measurement $\{\langle \gamma^{\mu | \nu}|\}$ given the axis $\mu$, where $b$ is the outcome of the site
measured in this basis, and $c$ is the outcome of the associated site (connected to
the former directly by an unmatched bond) measured in the standard basis of the axis $\nu$ .}
\label{byproduct}
\end{table}

\subsection{Two-qubit logical gate: CNOT}

Together with the arbitrary single-qubit logical gates on every logical wire, it is widely known that any 
entangling two-qubit logical gate is sufficient to achieve the universality of quantum computation
when the quantum circuit is composed from the set of these gates.
Here we take the target two-qubit logical gate to be the controlled-not (CNOT) gate,
$|0^z \rangle\langle 0^z | \otimes \openone + |1^z\rangle \langle 1^z| \otimes X $.

The two-qubit logical gate requires connecting a pair of the {\it degree-3} backbone sites with 
the $\top$ and $\bot$ bonds for each.
In particular, the CNOT uses the degree-3 $\top$ site with the axis $\mu_1 = z$ and the degree-3 $\bot$ site 
with $\mu_2 = x$.
One could in principle use a matched-bond cluster with the same axis, instead of the single degree-3 site.
Observe that
\begin{align}\begin{split}
A_{\top}[\gamma^{z|x} (0)] &= \tfrac{1}{\sqrt{2}} \Upsilon_{\top}
\left(\openone \otimes \bra{0^{x}} + Z \otimes \bra{1^{x}}\right),\\
A_{\bot}[\gamma^{x|z} (0)] &= \tfrac{1}{\sqrt{2}} \Upsilon_{\bot}
\left(\openone \otimes \ket{0^{z}} + X \otimes \ket{1^{z}}\right),
\label{cnot}
\end{split}\end{align}
where the byproducts are $\Upsilon_{\top} = XZ^{b_{\top}} \otimes \openone$ and
$\Upsilon_{\bot} = X^{b_{\bot} \oplus 1} Z \otimes Z$, and the conditioning axes $\nu_1 , \nu_2$ of the complementary 
bases can be chosen freely, so that we have set $\nu_1 = x$ and $\nu_2 = z$ for convenience.
So, when the vertical direction (the second degree of freedom in the tensor structure of Eq.~(\ref{cnot}))
is contracted through the sequence of the logical identities in the same way as in the part of a quantum logical wire,
it is readily seen that this provides the desired CNOT $(= \tfrac{1}{2}\left(\openone \otimes \openone + 
\openone \otimes X + Z \otimes \openone - Z\otimes X \right))$.

Here we need to calculate the total byproduct operator from this intermediate vertical part, namely
$\Upsilon_{|} = X^{\sum_{k} a^x_{k}} Z^{\sum_{k} a^z_{k}}$ where the summation is taken over 
the degree-2 backbone sites symbolically labeled by $k$.
Needless to say, if there are, in between, the branches by the matched bonds, we have to take
their contribution into account according to the prescription of the~\ref{app:match}, in order to 
determine $(a^x_{k}, a^z_{k})$ at every junction to the branch. Thus,
$\Upsilon_{|}$ essentially amounts to the whole accumulation of the byproducts from the vertical part.
By absorbing $Z^{\sum_k a^z_k \oplus 1}$ (which has included the additional one from $\Upsilon_{\bot}$) 
into $\bra{0/1^x}$ of $A_{\top}$, as well as $X^{\sum_k a^x_k}$ into $\ket{0/1^z}$ of $A_{\bot}$, 
we can reduce them to the byproduct operators {\it along two logical wires}.
The whole logical action is $\Upsilon {\rm CNOT}$, where 
$\Upsilon = XZ^{ b_{\top} \oplus \sum_k a^z_k \oplus 1} \otimes X^{ b_{\bot} \oplus \sum_k a^x_k \oplus 1} Z$.

\subsection{Initialization and readout}

The initialization and readout (in the sense of the quantum circuit model) of the logical wire can be simulated as well.
Suppose, without loss of generality, these are always made in the $\ket{0/1^z}$ basis.
All we have to do is to find the degree-2 backbone site with $\mu = z$ on the rightmost (for the initialization) and 
on the leftmost (for the readout) of every logical wire, and to measure it in the standard basis 
$\{\bra{\tfrac{3}{2}^z}, \bra{-\tfrac{3}{2}^z} \}$.
As already described through the notation for the standard-basis measurement, according to Eqs.~(\ref{Atop}, \ref{Abot}),
the outcome $c$ must be interpreted as $\ket{c^{z}}$ or $\bra{c \oplus 1^{z}}$.
Note, in particular regarding the readout, that the outcome $c$ at the leftmost boundary site is {\it not} immediately
the {\it logical} outcome of the computation. The latter is evaluated only together with the byproduct operator
at the end of the computation.

\subsection{Adaptation based on byproduct operators}

Since the byproduct operator $\Upsilon$ stays in the Pauli group, we can use the same machinery 
as the cluster-state MQC, to deal with the randomness of the measurement outcome.
The key idea is to postpone the effect of $\Upsilon$ by adapting the following angles of the logical rotations.
As an illustration, suppose we wish to apply the sequence of $R^z (\theta^z)$ followed by $R^x (\theta^x)$ up to
some byproduct operator $\Upsilon$, namely $ \Upsilon R^x (\theta^x) R^z (\theta^z)$.
Since
\beq
(X^{b' \oplus c'} Z R^x (\theta^x)) ( X Z^{b \oplus c} R^z (\theta^z)) =  
X^{b' \oplus c' \oplus 1} Z^{b \oplus c \oplus 1} R^x ((-1)^{b \oplus c} \theta^x) R^z (\theta^z) ,
\eeq
we realize that if we adapt the second angle $\theta^x$ to be $(-1)^{a^z} \theta^x$, based on the
byproduct index $a^z = b \oplus c$ of the first measurement, we can always apply the desired sequence
of the rotations regardless of the outcomes $b$ and $c$. 
In general, we have to adapt the angle of the next logical rotation, based on the current byproduct
operator $\Upsilon$ updated according to the simulated time direction of the quantum circuit.  
Note that such an adaptation is required only if the rotation angle is not fiducial, so that the part by the logical 
identities does not need the adaption and can be implemented parallely in principle.
Since this adaptation is now a widely-known machinery in MQC, we leave the details to 
the literatures, e.g., Ref.~\cite{raussendorf03} (although in the case of the cluster state there is an additional 
automatic logical operation by the Hadamard matrix $\tfrac{1}{\sqrt{2}}(X + Z)$ at every single step of 
computation).

\section{Global nature of emergent computational capability}
\label{sec:global}

\subsection{Identification of the backbone}

In this section, we show that the backbone structure, which should be intact from rare ``off-limits'' configurations of 
matched bonds, can be identified efficiently in analyzing their occurrence by classical side-computation. 
The analysis not only guarantees that the widgets for the logical gates in the Section~\ref{sec:gates}
can be composed consistently, but also suggests how the computational 
capability as a whole is linked to the emergence of a macroscopic feature of the many-body system.

Recall that, at the first stage of the protocol, every outcome of the polarizing measurement 
$\{M^{\mu}\}  \; (\mu = x,y,z)$ occurs randomly at every site, i.e., with the equal probability $\tfrac{1}{3}$.
Thus, in neglecting two-point correlation functions exponentially decaying in distance as a first approximation,
we can imagine that the bond between a pair of the neighboring sites is unmatched with probability 
$\tfrac{2}{3}$ and matched with probability $\tfrac{1}{3}$.
Indeed, the effect of two-point correlations is very limited, and it seems to modify the probability of the 
unmatched bond by not more than 1 percent. 
Accordingly, we could visualize in our mind a typical configuration of unmatched bonds, as
presented in the Figure~\ref{fig:matchedbonds}, by the proximity to  
the bond percolation model with an occupation probability $p=\tfrac{2}{3}$.
As a reference, it is widely known in the bond percolation model that, when $p$ is larger than the critical value 
$p_c = 1 - 2 \sin (\tfrac{\pi}{18}) \approx 0.652 \ldots $ \cite{sykes64}
for the 2D hexagonal lattice, there exists almost surely (i.e., with a probability close to the unity exponentially 
in the lattice size) a single macroscopic cluster of the occupied bonds, spanning the whole lattice.

Importantly, it is shown (in detail in the~\ref{app:match}) that not only unmatched bonds, which are
the majority, but also most of matched bonds are indeed available as a part of the backbone under our prescription.
The key idea is a ``topological encoding'' of the backbone structure 
in that in most cases the matched-bond cluster touching the backbone and measured in its 
complementary basis can be made to constitute a micro-circuit whose output to the backbone is 
mutually-unbiased as if the whole cluster were a single unmatched bond.
In other words, the matched-bond cluster can be effectively renormalized to the {\it single logical} site 
directly on the backbone (which is called the ``root'' site in the~\ref{app:match}), regardless of 
the size and shape of the cluster.
This prescription clarifies that the backbone should be identified in order that
the region measured in the complementary bases (namely, the backbone and all the matched-bond cluster touching it)
does not induce any additional closed loop which is absent in the target quantum circuit.

For a practical algorithmic implementation, we could leverage a variety of polynomial-time
classical algorithms (see e.g., Ref.~\cite{optimization}) developed through the study of percolation models.
For instance, using an existing method to detect the locations of matched-bond clusters locally from 
the records of the polarizing measurement outcome at every site, 
we can efficiently mark potential {\it off-limits} configurations that two matched-bond clusters with different axes 
are nearest neighboring via more than one unmatched bond (see the~\ref{app:match} for their precise notion 
related to the existence of closed loops).
This is because measuring both clusters in complementary bases must result in an undesired closed loop inevitably.
So, one could consider one of such clusters to be effectively ``unoccupied'' bonds (which are marked by 
thicker, light-purple bonds in the Figure~\ref{fig:matchedbonds}), and choose the backbone from the remaining 
(unmatched and matched) bonds.

Thus, our situation is much more favorable to find every horizontally percolating path corresponding to a quantum 
logical wire, compared to the case where such a path should be found from the cluster of unmatched bonds only.
The ratio of ``occupied'' bonds is much larger than $\tfrac{2}{3}$ effectively, and our macroscopic cluster of 
the occupied bonds can indeed offer abundant choices.
For this purpose, it is straightforward to apply to our context the same efficient classical algorithm to find multiple 
percolating paths, despite that the unoccupied bonds are not generated according to the standard percolation model.
That may involve, in practice, some customization such as a preference to the sites with certain axes
(depending on the choice of the elementary gates) and optimization about physical overheads, but
we do not intend to pursue concrete algorithms further.
Note that the use of a percolation phenomenon has been previously considered in the context of MQC but 
in rather different scenarios \cite{kieling07,browne08} where the preparation of a cluster state suffers 
some locally probabilistic errors.
For example, Ref.~\cite{browne08} has presented a detailed algorithm to extract explicitly
multiple percolating paths for a 2D cluster state. Such machinery to the cluster state is also used
in another approach to indicate computational usefulness of the 2D AKLT state in Ref.~\cite{wei11}.

Only a practically notable constraint by the matched bonds is on the spatial overhead of the backbone,
in particular, the spacing between every neighboring pair of quantum logical wires running horizontally.
Two quantum logical wires are not supposed to enter a single matched-bond cluster, since the latter
behaves as the single logical site effectively.
According to the approximation by the percolation model, it is expected that the size of the largest 
matched-bond cluster scales logarithmically in the lattice size but it occurs with an exponentially small probability.
Since we would likely resort to quantum error correction to the goal of a scalable fault-torelant quantum computer in the end,
we can set a cutoff within its constant error threshold, so as to disregard the rare large clusters and to deal with them 
in the similar way with other errors.
It means that a constant spacing between two neighboring logical wires would be sufficient.

\subsection{Emergence of a simulated space-time}

It is remarkable to see how the distinction between (simulated) space and time
emerges in our model.
Our backbone structure after the first stage does not distinguish them yet, but we can realize, by a careful 
examination of the protocol, it originates from the way the byproduct indices $(a^x, a^z)$, namely
two bits of classical information processed at each site, are communicated to the neighbors. 
Time emerges when both two indices are forwarded in the {\it same} (and roughly horizontal here) direction, 
while space emerges when two are sent in the {\it opposite} (and roughly vertical here) directions as is the case of 
CNOT, where $a^x$ or $a^z$ is sent downward or upward, respectively.
Note also that since the spatial-entangling gate, like CNOT, is simulated only using the 
measurements with the fiducial angle and thus need not be adapted, 
there is indeed no ``time-ordering'' along the simulated spatial direction.

\section{Discussion}
\label{sec:conclude}

\subsection{Ground-code MQC}

A feature that our resource is available as the ground state can be leveraged during
the computational process in addition to the stage of its preparation.
As originally proposed in Ref.~\cite{brennen08} as the ground-code version of MQC using 
coupled 1D AKLT states, if we are capable of adiabatically turning off a two-body interacting term 
only to the spin to be measured selectively --- importantly that still does not have 
to create new entanglement --- we could indeed make the Hamiltonian in the remaining bulk coexist 
without interfering the MQC protocol for quantum computation.
Then, the Hamiltonian with a (conjectured) gap would be able to contribute to a passive 
protection of the logical information stored in the degenerate ground state, by providing 
some robustness against local noises, in compared to a scenario where the bulk Hamiltonian is absent.

Here we need additional care to apply the idea of the ground code to the 2D AKLT state, since 
it was assumed that the spin $\tfrac{3}{2}$'s are first assigned 
to one of three axes through the polarizing measurement before the quantum computation starts. 
It turns out that it is still possible for the the classical algorithms described in the 
Section~\ref{sec:global} to function if the sampling of the matched bonds are made 
over the block of a constant number of spins near the rightmost boundary of the unmeasured bulk,
without identifying the full configuration of matched bonds deep inside the bulk.

\subsection{Persistence of computational capability in the VBS phase}

Another significant merit to have the bulk Hamiltonian during the computation
is that {\it any} ground state which belongs to the 2D VBS phase 
(with the rotational symmetry as is present in the AKLT Hamiltonian) is conceived 
to be ubiquitously useful to simulate the same quantum computation through the aforementioned 
adiabatic turning off of the interaction at the boundary. 
Its detail will be analyzed elsewhere, but such persistence of computational capability 
over an entire quantum phase was unveiled in Ref.~\cite{miyake10}, in examining the 
rotationally-invariant 1D VBS phase to which the 1D AKLT state belongs.
A key of the mechanism is indeed the edge states --- fractionalized degrees of freedom 
emergent at the boundary of the unmeasured bulk part --- that are commonly present
in the VBS phase and carry the logical information of quantum computation in our context.
At any point within the VBS phase, the quantum correlation between the spin to which the two-body interaction
has been turned off and the emergent edge state at a new boundary of 
the bulk would be modified to exactly that of the AKLT state, since the AKLT Hamiltonian satisfies
additionally a frustration-free property (that the global ground state also minimizes 
every summand of the Hamiltonian) among a parameterized class of generally-frustrated 
Hamiltonians within the phase. That is how the primitive of the ground-code MQC, the adiabatic turning off
of the interaction followed by the measurement of the freed spin, can ubiquitously unlock
computational capability persistent over the VBS phase.

In contrast, suppose our ability of the selective control of a single spin at the boundary is 
further limited only to its measurement and the bulk Hamiltonian is neither engineered nor present during the
computational process. We may still consider implementing physically (namely probabilistically in
compensation with a longer distance scale) a quantum computational renormalization, through which 
the correlation within the VBS phase is modified into that of a fixed point.
This is envisioned based on Ref.~\cite{bartlett10} where a protocol within the 1D VBS phase was constructed 
so as to set the AKLT point as its fixed point.
An attempt to construct a 2D counterpart of the protocol may face the intrinsic difficulty for the traditional 
computer to analyze correlations in the 2D system, but recent various developments of the classical methods 
(cf. in Refs.~\cite{vidal08,levin07,cirac09,gu09}) regarding the renormalization of 2D entanglement 
might be applicable to our context.

\subsection{Outlook towards physical implementations}

The persistence of such computational capability over the 2D VBS phase would be encouraging towards physical 
implementations of our scheme. Then a fine tuning to engineer a set of specific interaction strengths 
in the 2D AKLT Hamiltonian of Eq.~(\ref{H}) may not be necessary as far as the system can be set within the 
``quantum computational'' phase in maintaining some key symmetries such as the rotational symmetry.

The biggest difference from the 1D case is the fact that the ordinary Heisenberg point (i.e. the 
ground state of the Hamiltonian without biquadratic and bicubic terms in Eq.~(\ref{H})) has been known 
to be N{\'e}el ordered \cite{AKLT88} and does not belong to the 2D VBS phase unfortunately, 
although the 1D Heisenberg point is located within the 1D VBS phase. 
That implies that we may need a more elaborated interaction than the simplest Heisenberg interaction
or a use of geometrically frustrated quantum magnets, in order to set a ground state in the 2D VBS phase. 
In fact, the 1D VBS phase has been identified experimentally in several natural chemical 
compounds such as \mbox{CsNiCl$_{3}$}, \mbox{Y$_{2}$BaNiO$_{5}$}, and so-called NENP.
The 2D VBS phase seems to be less studied both theoretically and experimentally, compared to 
other kinds of 2D VBC phases for now. A recent discovery of the absence of spontaneous magnetic
ordering in the manganese oxide \mbox{Bi$_{3}$Mn$_{4}$O$_{12}$(NO$_{3}$)} \cite{smirnova09}, which materializes
an antiferromagnet of spin $\tfrac{3}{2}$'s in terms of \mbox{Mn$^{4+}$} ions on the 2D hexagonal lattice, 
might pave the way towards this direction (e.g., Ref.~\cite{ganesh10}).

Another possible realization of the 2D VBS phase is based on analog engineering of the Hamiltonians of 
spin lattice systems.
A recent experimental development of quantum simulators in atomic, molecular, and optical physics 
is promising in that Mott insulating phases of ultracold Fermi spinor gases with a hyperfine manifold 
$F=\tfrac{3}{2}$, e.g., in terms of $^{6}$Li or $^{132}$Cs atoms trapped in the optical lattice, might be 
used to simulate the 2D AKLT Hamiltonian of spin $\tfrac{3}{2}$'s along an extension of the approach in 
Ref.~\cite{eckert07} for example.

\subsection*{Acknowledgments}

I acknowledge various relevant discussions, over a long period of the project, with 
I.~Affleck, S.D.~Bartlett, G.K.~Brennen, H.J.~Briegel, J.-M. Cai, W.~D\"{u}r, D.~Gross, M.B.~Hastings,
H.~Katsura, R.~Raussendorf, J.M.~Renes, A.~Sanpera, N. Schuch, and T.-C.~Wei.
The work is supported by the Government of Canada through Industry Canada and by Ontario-MRI.

\appendix

\section{Boundary tensor in the 2D VBS state}
\label{app:boundary}

The boundary tensor $B$ is defined as $B = \prod_{\ell} \Upsilon_{\ell}$, where $\ell$ represents a nonlocal degree
of freedom associated with every {\it pair} of the boundary sites which are supposed to be closed under the periodic 
boundary condition, and thus $2 | \ell |$ represents the total number of the boundary sites.
This term formally provides the $4^{ | \ell |}$-fold initial degeneracy to the ground state, but it turns out that 
any of such ground states (and actually even the mixed state by them through an argument similar to Ref.~\cite{miyake10}) 
is useful to our goal, because of the decoupling property to the backbone. 
Thus, here we set $B$ to be the identity for simplicity.

\section{Renormalizing prescription to matched bonds}
\label{app:match}

We elaborate a technical prescription for matched bonds, and prove the following informal lemma.

\begin{lemma}
The backbone is freely chosen as far as the global topology of the region measured in the complementary bases 
(namely, the backbone and all the matched-bond clusters touching to it) is equivalent to the target quantum circuit,
regarding the configuration of closed loops.
\end{lemma}

The lemma can be also seen as our resolution to the genuine 2D fluctuations present in naturally-occurring systems.
They are dealt, by adjusting the encoding of each single logical site to be {\it variable} according to the size
and shape of every matched-bond cluster.

\subsection{1D-like chain}

Let us consider first a 1D-like chain of the matched bonds,
branching out from a degree-2 site on the backbone of the axis $\mu$.
Here we call such a site as the {\it root} site of the matched-bond cluster.
The sequence does not bifurcate further, and thus is expected to terminate in the end 
at a non-backbone site after a chain of matched bonds. 
Let us for now label these sites along the matched-bond chain by $k$.
As said, all the non-backbone sites along the chain of the matched bonds are measured in the complementary
bases $\{\bra{\gamma^{\mu | \nu_{k}}}\}$, which depend on the axis $\nu_k$ of each associated site connected 
by the {\it unmatched} bond in the third direction orthogonal to the matched-bond chain.
At the end of the matched-bond chain, the terminating site must have two unmatched bonds, and its
two associated sites are measured in the standard basis of each.
That provides a pair $\ket{c^{\nu}}$ and $| c'^{\nu'}\rangle$ to the tensors at the terminating site, so that either one,
here $| c'^{\nu'}\rangle$, can be used to assign $\tilde{A}[\alpha]$ to the terminating site as usual, 
and the other $\ket{c^{\nu}}$ plays a role of the ``initial'' condition to the matched-bond chain.
In the exactly same way as the chain of the degree-2 backbone sites, this matched-bond chain 
behaves as the sequence of the logical identities acting on the initial state $\ket{c^{\nu}}$.
At the root site to the backbone, the total logical effect is equivalent to input $\ket{\bar{c}^{\bar{\nu}}}$
given by
\beq
\ket{\bar{c}^{\bar{\nu}}}  =\!\!  \prod_{k}^{\rm matched} \!\! \tilde{A}_{k}[\gamma^{\mu | {\nu_k}} (0)] |c^{\nu}\rangle 
= X^{\sum_{k} a^{x}_k} Z^{\sum_{k} a^{z}_k} |c^{\nu}\rangle ,
\label{branch}
\eeq
where $\bar{\nu} = \nu$ and
\beq
\bar{c} = \left\{\begin{array}{ll}
 c \oplus \sum_{k} a^z_k & \mbox{if} \;\; \nu = x , \\
 c \oplus \sum_{k} (a^x_k \oplus a^z_k) & \mbox{if} \;\;   \nu = y , \\
 c \oplus\sum_{k} a^x_k   & \mbox{if} \;\; \nu = z .
\end{array} 
\right.
\label{barc}
\eeq
The indices $(a^{x}_{k}, a^{z}_{k})$ of the byproduct operator at every site $k$, measured in the complementary 
basis along the matched-bond chain, are determined according to the rule at the Table~\ref{byproduct} (in
incorporating the contributions from associated sites measured in the standard basis). 
$\bra{\bar{c}\oplus 1^{\bar{\nu}}}$ is determined exactly in the same formula.
The key observation is that the matched-bond chain all measured in the complementary basis provides 
a ``mutually-unbiased'' input state in an axis $\nu$ different from the axis $\mu$ of the matched bonds, and 
thus can play a comparable role to a single unmatched bond.

\subsection{Locally tree structure}

In general, the branching-out cluster of the matched bonds may have bifurcations like a tree locally.
Flowing backward from every leaf of the tree (or, the terminating site of a matched-bond chain) according to the 
aforementioned procedure, we can prepare the mutually-unbiased input at each bifurcation site.
Thus, we concatenate this procedure several times until it reaches the root site of 
the matched-bond cluster located on the backbone.

\subsection{Closed loop}

For completeness, we remark that closed loops may exist within the matched-bond cluster, branching
out from the backbone.
It essentially corresponds to the counterpart of Eq.~(\ref{branch}) in a closed boundary condition.
Suppose a closed loop, for example a hexagon, exists when we are applying the aforementioned
procedure from every leaf of the cluster. 
Each site in the closed loop, except the the one which is closest to the root site of the cluster
and called as the zeroth here, is now assumed to be associated effectively with 
the mutually-unbiased input using either an unmatched bond or our prescription for matched bonds.
Then, the contribution by the closed loop is 
\beq
{\rm tr}_{\rm \, loop} \left[A_{0}[\gamma^{\mu | {\nu_0}} (0)] \!\! \prod_{k}^{\rm matched} \!\! 
\tilde{A}_{k}[\gamma^{\mu | {\nu_k}} (0)] \right] ,
\label{loop}
\eeq
where the axis $\nu_0$ can be chosen to be either axis different from $\mu$,
and note that $A_0$ is the original, three-directional tensor.
For example, if $\mu = z$ and the zeroth is a $\top$ site, Eq.~(\ref{loop}) is equivalent to
\beq
\sum_{\bar{c}=0,1} 
{\rm tr} \left[ X^{1 \oplus \sum_k a^{x}_k} Z^{b_0 \oplus \bar{c} \oplus 1 \oplus \sum_k a^{z}_k}\right] 
\otimes \bra{\bar{c}\oplus 1^{\nu_0}} ,
\eeq
by denoting the zeroth-site bare outcome as $b_0$.
Remarkably, careful examination of the Table~\ref{byproduct} clarifies that 
the total byproduct operator along the loop could accumulate up to the identity operator 
(with a regular weight $\tfrac{1}{\sqrt{2}}$), and the contribution of Eq.~(\ref{loop}) is always non-vanishing.
In this example, $1 \oplus \sum_k a^{x}_k \equiv 0$ because the number of sites constituting a loop 
must be {\it even} and every $a^{x}_k$ is always $1$ according to the Table~\ref{byproduct}.
Thus, the closed loop of matched bonds provides likewise
a mutually-unbiased input $\ket{\bar{c}^{\bar{\nu}}}$ or $\bra{\bar{c}\oplus 1^{\bar{\nu}}}$
towards the root site, where $\bar{\nu} = \nu_0$ and
\beq
\bar{c} \oplus 1 = \left\{\begin{array}{ll}
 b_0 \oplus \sum_{k} a^x_k & \mbox{if} \;\; \mu = x , \\
 b_0 \oplus \sum_{k} a^x_k  =  b_0 \oplus \sum_{k} (a^z_k \oplus 1) & \mbox{if} \;\;   \mu = y , \\
 b_0 \oplus \sum_{k} a^z_k   & \mbox{if} \;\; \mu = z ,
\end{array} 
\right.
\eeq
in analogy to Eq.~(\ref{barc}).

This analysis suggests that the closed loop itself can be directly a part of the backbone, namely can share one or more 
bonds with the backbone.
Phrased differently, the matched-bond loop (of the same axis) does not disturb the global topology of the backbone.
This can be confirmed by inserting, instead of $A_0$, {\it two} original three-directional tensors, e.g. $A_{\rm in}$ 
and  $A_{\rm out}$ at a $\top$ and a $\bot$ site respectively with $\mu = z$ and $\nu_0 = x$ or $y$, in Eq.~(\ref{loop}). 
Note that the two are not necessarily the nearest neighbors, nor necessarily $\top$ and $\bot$ sites in general.
\begin{align}
& {\rm tr}_{\rm \, loop} \left[A_{\rm out}[\gamma^{z | {\nu_0}} (0)] \!\! \prod_{k}^{\rm matched} \!\! 
\tilde{A}_{k}[\gamma^{z | {\nu_k}} (0)]  A_{\rm in}[\gamma^{z | {\nu_0}} (0)] \right]  \nonumber\\
& = \sum_{c, c' =0,1} \!\!
{\rm tr} \left[X^{2 \oplus \sum_k a^{x}_k} Z^{b_{\bot} \oplus c' \oplus \sum_k a^{z}_k  \oplus b_{\top} \oplus c} \right] 
\otimes |{c'}^{\nu_0}\rangle \bra{c \oplus 1^{\nu_0}}
= Z^{b_{\top} \oplus b_{\bot} \oplus \sum_{k} a^{z}_k \oplus 1} ,
\label{loop2}
\end{align}
where in the last equality we have used the non-vanishing condition of the trace,
$c \oplus c' = b_{\top} \oplus b_{\bot} \oplus \sum_{k} a^{z}_k$ (i.e., the parity of $c$ and $c'$ is fixed),
based on the property that the exponent of the Pauli $X$ operator is identically zero once again.
Thus, the contribution by the loop is a power of the Pauli $\mu$ operator depending on the measurement
outcomes in an extended manner of the Table~\ref{byproduct}. It is also straightforward to see that
we could get further $R^z (\theta)$ by this matched-bond cluster, by replacing one of fiducial rotations by $\theta$, as usual.

In a nutshell, as mentioned in the main text, the cluster of matched bonds can be effectively 
treated as a single logical site under our renormalizing prescription, regardless of its size and shape.

\subsection{Off-limits configuration}

On the contrary, suppose a closed loop of sites measured in the complementary bases
is introduced by a situation that the backbone touches sequentially
two kinds of matched-bond clusters (with different axes, say $\mu =x$ and $z$) nearest neighboring 
via two or more unmatched bonds. 
Then the contribution of the induced loop does change the global topology of the backbone.
As an illustration, suppose the induced loop consists of the matched-bond cluster of the $z$ axis and a single 
site of the $x$ axis, all of which are measured in their own complementary bases. 
By replacing in Eq.~(\ref{loop2}) the axis $z$ at $A_{\rm in}$ by $x$ (implicitly there are two unmatched bonds in the 
loop), we see that
\begin{align}
& {\rm tr}_{\rm \, loop} \left[A_{\rm out}[\gamma^{z | {\nu'_0}} (0)] 
\!\! \prod_{k}^{\rm matched} \!\!  \tilde{A}_{k}[\gamma^{z | {\nu_k}} (0)]  A_{\rm in}[\gamma^{x | {\nu_0}} (0)] \right]  \nonumber\\
& = \sum_{c, c' =0,1} \!\!
{\rm tr} \left[X^{b_{\top} \oplus c \oplus \sum_k a^{x}_k} Z^{b_{\bot} \oplus c' \oplus \sum_k a^{z}_k} \right] 
\otimes |c'^{\nu'_0}\rangle \bra{c \oplus 1^{\nu_0}} 
= |b_{\bot} \oplus \mbox{$\sum_k$} a^{z \, \nu'_0}_k\rangle \bra{b_{\top} \oplus 1^{\nu_0}} .
\end{align}
The resulting map is not unitary anymore in contrast to Eq.~(\ref{loop2}).
Such an off-limits configuration would be readily circumvented through local analysis of the configuration of 
matched-bond clusters.  
In the Figure~\ref{fig:matchedbonds}, we have illustrated marking (by light purple) the off-limits configurations,
where each cluster is paired with another cluster in such a manner that both should not be measured 
in the complementary bases.



\begin{thebibliography}{99}

\bibitem{divincenzo00}
D.P. DiVincenzo, Fortschr. Phys. {\bf 48}, 771 (2000).

\bibitem{feynman82}
R. Feynman, Int. J. Theor. Phys. {\bf 21}, 467 (1982).

\bibitem{AKLT87}
I. Affleck, T. Kennedy, E.H. Lieb, and H. Tasaki, Phys. Rev. Lett. {\bf 59}, 799 (1987).

\bibitem{AKLT88}
I. Affleck, T. Kennedy, E.H. Lieb, and H. Tasaki,  Comm. Math. Phys. {\bf 115}, 477 (1988).

\bibitem{briegel01}
H.J. Briegel and R. Raussendorf, Phys. Rev. Lett. {\bf 86}, 910 (2001).

\bibitem{raussendorf01}
R. Raussendorf and H.J. Briegel, Phys. Rev. Lett. {\bf 86}, 5188 (2001).

\bibitem{raussendorf03}
R. Raussendorf, D.E. Browne, and H.J. Briegel, Phys. Rev. A {\bf 68}, 022312 (2003).

\bibitem{verstraete04}
F. Verstraete and J.I. Cirac, Phys. Rev. A {\bf 70}, 060302(R) (2004).

\bibitem{gross07prl}
D. Gross and J. Eisert, Phys. Rev. Lett. {\bf 98}, 220503 (2007).

\bibitem{gross07pra}
D. Gross, J. Eisert, N. Schuch, and D. Perez-Garcia, Phys. Rev. A {\bf 76}, 052315 (2007).

\bibitem{brennen08}
G.K. Brennen and A. Miyake, Phys. Rev. Lett. {\bf 101}, 010502 (2008).

\bibitem{gross08} 
D. Gross and J. Eisert, Phys. Rev. A {\bf 82}, 040303(R) (2010).

\bibitem{bartlett06}
S.D. Bartlett and T. Rudolph, Phys. Rev. A {\bf 74}, 040302(R) (2006).

\bibitem{griffin08}
T. Griffin and S.D. Bartlett, Phys. Rev. A {\bf 78}, 062306 (2008).

\bibitem{chen09}
X. Chen, B. Zeng, Z.-C. Gu, B. Yoshida, and I.L. Chuang, Phys. Rev. Lett. {\bf 102}, 220501 (2009).

\bibitem{cai10} 
J.-M. Cai, A. Miyake, W. D\"{u}r, and H.J. Briegel, Phys. Rev. A {\bf 82}, 052309 (2010).

\bibitem{nielsen06}
M.A. Nielsen, Rep. Math. Phys. {\bf 57}, 147 (2006).

\bibitem{nest08}
M. Van den Nest, K. Luttmer, W. D\"{u}r, and H.J. Briegel, Phys. Rev. A {\bf 77}, 012301 (2008)

\bibitem{wei11} 
T.-C. Wei, I. Affleck, and R. Raussendorf, Phys. Rev. Lett. {\bf 106}, 070501 (2011).

\bibitem{arovas88}
D.P. Arovas, A. Auerbach, and F.D.M. Haldane, Phys. Rev. Lett. {\bf 60}, 531 (1988).

\bibitem{KLT88}
T. Kennedy, E.H. Lieb, and H. Tasaki, J. Stat. Phys. {\bf 53}, 383 (1988).

\bibitem{katsura10} 
H. Katsura, N. Kawashima, A.N. Kirillov, V.E. Korepin, and S. Tanaka, J. Phys. A: Math. Theor. {\bf 43}, 255303 (2010).

\bibitem{sykes64}  
M.F. Sykes and J.W. Essam, J. Math. Phys. {\bf 5}, 1117 (1964).

\bibitem{optimization}
A.K. Hartmann and H. Rieger, ``Optimization algorithms in physics,'' (Wiley-VCH, Berlin, 2001).

\bibitem{kieling07}
K. Kieling, T. Rudolph, and J. Eisert, Phys. Rev. Lett. {\bf 99}, 130501 (2007).

\bibitem{browne08}
D.E. Browne, M.B. Elliott, S.T. Flammia, S.T. Merkel, A. Miyake, and  A.J. Short, New J. Phys. {\bf 10}, 023010 (2008).

\bibitem{miyake10}
A. Miyake, Phys. Rev. Lett. {\bf 105}, 040501 (2010).

\bibitem{bartlett10}
S.D. Bartlett, G.K. Brennen, A. Miyake, and J.M. Renes, Phys. Rev. Lett. {\bf 105}, 110502 (2010).

\bibitem{vidal08}
G. Vidal, Phys. Rev. Lett. {\bf 101}, 110501 (2008).

\bibitem{levin07}
M. Levin and C.P. Nave, Phys. Rev. Lett. {\bf 99}, 120601 (2007).

\bibitem{cirac09}
J.I. Cirac and F. Verstraete, J. Phys. A: Math. Theor. {\bf 42}, 504004 (2009).

\bibitem{gu09}
Z.-C. Gu and X.-G. Wen, Phys. Rev. B {\bf 80}, 155131 (2009).

\bibitem{smirnova09}
O. Smirnova, M. Azuma, N. Kumada, Y. Kusano, M. Matsuda, Y. Shimakawa, T. Takei, Y. Yonesaki, and N. Kinomura, 
J. Am. Chem. Soc. {\bf 131}, 8313 (2009).

\bibitem{ganesh10} 
R. Ganesh, D.N. Sheng, Y.-J. Kim, and A. Paramekanti, Phys. Rev. B {\bf 83}, 144414 (2011).

\bibitem{eckert07}
K. Eckert, {\L}. Zawitkowski, M.J. Leskinen, A. Sanpera, and M. Lewenstein, New J. Phys. {\bf 9}, 133 (2007).

\end{thebibliography}
\end{document}